\documentclass[10pt]{article}
\usepackage{bbm}
\usepackage[final]{graphics}
\usepackage{amsmath}
\usepackage{amsfonts,amsbsy}
\usepackage{amssymb}
\usepackage{latexsym}
\usepackage[latin1]{inputenc}
\usepackage[dvips]{hyperref}

\def\empile#1\over#2{\mathrel{\mathop{\kern 0pt#1}\limits_{#2}}}

\newcommand{\as}{{\alpha_{\rm s}}}

\newcommand{\xt}{{\boldsymbol{x}_\perp}}

\newcommand{\ut}{{\boldsymbol{u}_\perp}}
\newcommand{\yt}{{\boldsymbol{y}_\perp}}

\newcommand{\pt}{{\boldsymbol{p}_\perp}}
\newcommand{\qt}{{\boldsymbol{q}_\perp}}

\newcommand{\nc}{{N_\mathrm{c}}}

\newcommand{\nr}[1]{(\ref{#1})} 
\newcommand{\ud}{\mathrm{d}}
\newcommand{\fig}{fig.~}

\newcommand{\eq}{eq.~}

\def\p{{\boldsymbol p}}
\def\q{{\boldsymbol q}}

\def\k{{\boldsymbol k}}

\def\x{{\boldsymbol x}}
\def\y{{\boldsymbol y}}

\def\v{{\boldsymbol v}}

\def\u{{\boldsymbol u}}

\def\bs{\boldsymbol}
\def\wt#1{\widetilde{#1}}

\catcode`\@=11

\newcount\@tempcntc
\def\@citex[#1]#2{\if@filesw\immediate\write\@auxout{\string\citation{#2}}\fi
  \@tempcnta\z@\@tempcntb\m@ne\def\@citea{}\@cite{%
        \@for\@citeb:=#2\do%
    {\@ifundefined{b@\@citeb}%
        {\@citeo\@tempcntb\m@ne\@citea%
                \def\@citea{,\penalty\@m\ }{\bf ?}\@warning%
                {Citation `\@citeb' on page \thepage \space undefined}}%
        {\setbox\z@\hbox{\global\@tempcntc0\csname b@\@citeb\endcsname\relax}
     \ifnum\@tempcntc=\z@ \@citeo\@tempcntb\m@ne%
       \@citea\def\@citea{,\penalty\@m}%
       \hbox{\csname b@\@citeb\endcsname}%
     \else%
      \advance\@tempcntb\@ne%
      \ifnum\@tempcntb=\@tempcntc%
      \else\advance\@tempcntb\m@ne\@citeo%
      \@tempcnta\@tempcntc\@tempcntb\@tempcntc\fi\fi}}\@citeo}{#1}}%

\def\@citeo{\ifnum\@tempcnta>\@tempcntb\else\@citea
  \def\@citea{,\penalty\@m}%
  \ifnum\@tempcnta=\@tempcntb\the\@tempcnta\else
   {\advance\@tempcnta\@ne\ifnum\@tempcnta=\@tempcntb \else
\def\@citea{--}\fi
    \advance\@tempcnta\m@ne\the\@tempcnta\@citea\the\@tempcntb}\fi\fi}

\catcode`\@=12

\begin{document}

\title{\bf High energy factorization\\ in nucleus-nucleus collisions\\
  III. Long range rapidity correlations}

\date{}

\author{Fran\c cois Gelis$^{(1,2)}$, Tuomas Lappi$^{(2)}$, Raju Venugopalan$^{(3)}$}
\maketitle
\begin{center}
\begin{enumerate}
\item Theory Division, PH-TH, Case C01600, CERN,\\
 CH-1211, Geneva 23, Switzerland
\item Institut de Physique Th\'eorique (URA 2306 du CNRS)\\
  CEA/DSM/Saclay, B\^at. 774\\
  91191, Gif-sur-Yvette Cedex, France
\item  Physics Department, Brookhaven National Laboratory\\
  Upton, NY-11973, USA
\end{enumerate}
\end{center}

\maketitle

\begin{abstract}
  We obtain a novel result in QCD for long range rapidity correlations
  between gluons produced in the collision of saturated high energy
  hadrons or nuclei. This result, obtained in a high energy
  factorization framework, provides strong justification for the
  Glasma flux tube picture of coherent strong color fields.  Our
  formalism can be applied to ``near side ridge'' events at RHIC and
  in future studies of long range rapidity correlations at the LHC.
\end{abstract}

\section{Introduction}
Long range rapidity correlations in high ener\-gy ha\-dro\-nic
collisions are of interest in QCD because causality dictates that
these correlations are produced at very early times. They therefore
provide insight into how color correlations in the hadron
wavefunctions become dynamically manifest in multiparticle final
states.  Recent observations in nucleus--nucleus collisions at RHIC of
a ``near side ridge'' structure in two-particle
correlations~\cite{Putschke:2007mi,Daugherity:2008su,Wenger:2008ts}
and significant forward-backward multiplicity
correlations~\cite{Srivastava:2007ei} have reinvigorated interest in
the underlying dynamics of these correlations in QCD.  At the LHC,
with its wider rapidity coverage, such correlation studies can prove a
powerful diagnostic tool both of multiparton correlations in QCD and
of highly coherent strong color fields generated at early times in
nuclear collisions.

Long range rapidity correlations were previously studied in color flux
tube models where the non--perturbative dynamics is at the QCD scale
$\Lambda_{_{\rm QCD}}\sim 1$ fm$^{-1}$. However, the rapid growth of
parton distributions, and the requirement that occupation numbers in
QCD saturate at $\sim 1/\as$, where $\as$ is the QCD coupling
constant, suggests that the dynamics of color correlations is
controlled instead by a semi-hard saturation scale $Q_s\gg
\Lambda_{_{\rm QCD}}$. The properties of gluons with maximal
occupation are described by the Color Glass Condensate
(CGC)~\cite{CGC}; this saturated regime and the approach to it can be
computed in a weak coupling effective field theory (EFT).

In the CGC EFT, partons with longitudinal momenta $k^+$ larger than a
cutoff momentum $\Lambda^+$ in a high energy hadron or nucleus (moving
in the $+z$ direction) are described as static light-cone color
sources while modes with $k^+<\Lambda^+$ are treated as QCD gauge
fields that couple to these color sources~\cite{MV}. Because the
physics is independent of this separation scale, one obtains a
renormalization group (RG) equation--the JIMWLK
equation~\cite{JIMWLK}--describing the evolution of the distribution
of fast sources as $\Lambda^+$ is lowered.

The QCD matter formed immediately after a nucleus-nucleus collision is
a Glasma~\cite{Lappi:2006fp}. At leading order (LO), solutions of the
Yang--Mills equations reveal that the Glasma corresponds at early
times $\tau\lesssim Q_s^{-1}$ to highly coherent longitudinal
chromoelectric and chromomagnetic field
configurations~\cite{Kovner:1995ts,Kharzeev:2001ev} with maximal
occupation numbers $1/\as$.  At leading order, the Glasma fields are
invariant under boosts in the space--time rapidity $\eta$ and only
depend on their transverse positions in the nuclei and the proper time
$\tau$.  Further, the LO Glasma fields have the spatial structure of
flux tubes stretching between the two nuclei, each localized
transversely in a region of size $1/Q_s$. This geometrical picture
naturally explains the rapidity correlations observed in the near side
ridge in heavy ion collisions~\cite{Ridge,Ridge1}. However, the boost
invariance of the Glasma configurations at LO is broken by quantum
effects and it is important to understand their impact on
multiparticle correlations.

In two previous papers \cite{Gelis:2008rw,Gelis:2008ad} --hereafter
referred to as papers I and II-- we applied this effective field
theory to the inclusive multigluon spectra in nucleus-nucleus
collisions. The main result in these papers is a proof of the fact
that all the leading logarithms that arise in loop corrections to
these quantities can be absorbed into universal distributions for the
fast sources of the two nuclei. However, in the case of the
multigluon spectra, our proof was limited to the very peculiar
situation where all the observed gluons lie in a small region in
rapidity (of size $\Delta y \ll \as^{-1}$). This limitation was due to
the fact that we did not resum corrections of the form $\as|y_i-y_j|$
where $y_{i,j}$ are the rapidities of the gluons $i$ and $j$. These
corrections become important when the rapidity separation between the
observed gluons is of order $\as^{-1}$ or larger. Physically, these
corrections arise from the radiation of extra gluons between those
that are measured. This has a high probability of occurring if the
rapidity interval between two measured gluons is larger than
$\as^{-1}$.

The goal of the present paper, the third in this series, is to extend
the treatment in the previous papers to compute inclusive multigluon
spectra (to leading logarithmic accuracy) for arbitrary rapidity
intervals between the observed gluons. In the case of the two-gluon
spectrum, this is the basis for a detailed quantitative study of long
range rapidity correlations in heavy ion collisions\footnote{In
  \cite{Ridge,Ridge1}, a simpler leading order formula was used that
  does not resum the corrections in $\as|y_i-y_j|$.  This was
  sufficient to justify the existence of long range rapidity
  correlations and to suggest its relevance for the RHIC data. In
  particular, for STAR data, where the relevant rapidity window is
  $\Delta y \sim 1.5$, these resummation effects are not likely to be
  large. Resummations of long range rapidity corrections are however
  expected to improve the quantitative description of the STAR and
  PHOBOS data~\cite{Daugherity:2008su,Wenger:2008ts} and in future of
  the LHC data that has a significantly wider rapidity coverage.}.
These results are new and are valid (for any number of colors
including the physical $N_c=3$) in a weak coupling scheme where higher
order in $\as$ contributions--enhanced by the same powers of the
rapidity--are resummed to all orders. In this leading logarithmic
approximation, we will demonstrate that expectation values of
operators can be factorized as a convolution of density functionals
from each of the nuclei times the operator computed with leading order
classical fields. These density functionals evolve according to the
JIMWLK equation and are universal; they can be extracted, for
instance, in electron-nucleus collisions. Albeit the focus here will
be on nucleus--nucleus collisions, the results also apply to the
collision of two ``saturated'' hadrons at very high energies.

We note that while multiparticle correlations in the strong
interactions have been extensively
studied~\cite{oldstuff,JalilianMarian:2004da}, none of the literature
addresses nucleus--nucleus collisions, for finite $N_c$, in a
framework where gluon fields are the dynamical degrees of freedom.  We
will comment later on interesting earlier studies~\cite{oldstuff} on
two-particle correlations in the context of Local Reggeon Field
Theory.

The essence of the ``technology'' needed to resum all the leading logs
in multigluon spectra for arbitrary rapidity separations is already
contained in papers I and II, albeit in a somewhat hidden form. In
section \ref{sec:review}, we review the main results of these papers
and we prove a general formula for the renormalization group flow in
the CGC when one moves the cutoff $\Lambda^+$ of the effective theory
by an infinitesimal amount. In section \ref{sec:1and2gluon-spectra},
we show how 1- and 2-gluon inclusive spectra can be obtained from this
master formula.  Our formula for the 2-gluon spectrum is expressed in
terms of the usual distributions of color sources, and of a new object
that has the interpretation of a propagator (in functional space) for
the JIMWLK evolution. In the limit where the two gluons are nearby in
rapidity, we recover known results for the 2-gluon spectrum.  Finally,
in section \ref{sec:dilute}, we discuss the complications that arise
when one wants to extend these results to the case where one (or both)
of the projectiles is dilute instead of being in the saturation
regime. We end with a brief summary and outlook for future work.

\section{Leading logarithms at NLO}
\label{sec:review}
\begin{figure}[htbp]
\begin{center}
\resizebox*{!}{3.1cm}{\includegraphics{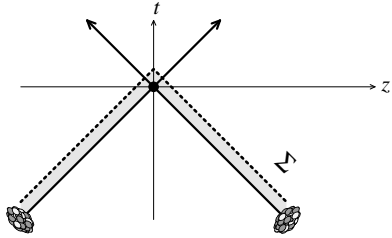}}
\end{center}
\caption{\label{fig:sigma}Initial space-time surface $\Sigma$ used in
  the initial value problem for the retarded classical field ${\cal
    A}^\mu$.}
\end{figure}

Consider an inclusive multigluon field operator ${\cal O}$. In
papers I and II, we have shown that its LO value ${\cal O}_{_{\rm
    LO}}$ can be expressed in terms of light-cone gauge classical
Yang--Mills solutions ${\cal A}^\mu$, with {\it retarded} boundary
conditions $\lim_{x^0\to-\infty} {\cal A}^\mu(x)=0$.  In coordinate
space, in the CGC effective field theory, the classical sources are
localized along the light-cones in two strips $0\le x^-\le
1/\Lambda^+$ and $0\le x^+\le 1/\Lambda^-$.  We denote by $\Sigma$ the
surface located at a distance $\delta x^\pm = 1/\Lambda^\mp$ above the
backward light-cone,
as illustrated in \fig \ref{fig:sigma}. Because the classical fields
involved in ${\cal O}_{_{\rm LO}}$ obey retarded boundary conditions,
${\cal O}_{_{\rm LO}}$ can be obtained by solving an initial value
problem with initial conditions defined on the surface $\Sigma$.

It is convenient to describe color sources in the nuclei by
distributions of Wilson lines 
\begin{equation}\label{eq:wilson-def}
\Omega_{1,2}(y,\x_\perp)\equiv
{\rm T}\,\exp ig \int_0^{x^\mp_y} dz^\mp\, 
\frac{1}{{\bs\nabla}_\perp^2}\wt{\rho}_{1,2}(z^\mp,\x_\perp)\; ,
\end{equation}
where ${\wt\rho}_{1,2}$ are the color source densities in Lorenz
gauge. Here the upper bound $x^\mp_y$ in the integral is related to
$y$ by $y\equiv\ln(P^\pm x^\mp_y)$, with $P^\pm$ the longitudinal
momenta of the respective nuclei~\footnote{With this convention, $y$
  is the rapidity measured from the beam rapidity and $y=0$
  corresponds to including only valence partons that have $0\le
  x^\mp\le 1/P^\pm$.}.

In papers I and II, the next-to-leading order (NLO) inclusive
multigluon spectrum was shown to take the form
\begin{eqnarray}
&&
{\cal O}_{_{\rm NLO}}
=
\left\{
\int\limits_{\Sigma}\ud^3\vec\u\,
\big[\beta\cdot{\mathbbm T}_\u\big] 
\right.
\nonumber\\
&&\qquad
+
\frac{1}{2}\sum_{\lambda,a}\int\frac{\ud^3\k}{(2\pi)^32E_\k}
\left.
\int\limits_{\Sigma} \ud^3\vec\u\, \ud^3\vec\v\,
\big[a_{-\k\lambda a}\cdot{\mathbbm T}_\u\big]
\big[a_{+\k\lambda a}\cdot{\mathbbm T}_\v\big]
\right\}
{\cal O}_{_{\rm LO}}\, ,
\label{eq:O-NLO}
\end{eqnarray}
up to terms that do not contribute at leading logarithmic accuracy.
The fields $\beta^\mu(x)$ and $a^\mu_{\pm\k\lambda a}(x)$ are small
quantum fluctuations propagating over the classical field ${\cal
  A}^\mu$.  The former has a vanishing boundary condition in the
remote past and its evolution is driven by a source term consisting of
a 1-loop tadpole graph.  The latter evolves without any source term,
but its boundary condition in the remote past is a plane wave $T^a
\epsilon_\lambda^\mu \,e^{\pm ik\cdot x}$ ($\lambda, a$ and $\k$
denote the polarization, color and momentum of the initial
fluctuation). Here, $d^3\vec\u$ is the measure on $\Sigma$.  Further, ${\cal
  O}_{_{\rm LO}}$ is a functional of the value of the classical field
${\cal A}^\mu$ on the surface $\Sigma$ and the differential operator
${\mathbbm T}_\u$ acting on ${\cal O}_{_{\rm LO}}$ is the generator of
shifts of the value of ${\cal A}^\mu$ on $\Sigma$.  The \emph{only}
part of the expression for this operator that is important for
computing leading logarithmic contributions is\footnote{The complete
  operator $a\cdot{\mathbbm T}_\u$ is made of three terms, but two of
  them (not written explicitly here) do not provide any leading
  logarithmic contributions.}
\begin{equation}\label{eq:aTu}
a\cdot{\mathbbm T}_\u
=
\partial_\mu(\Omega_{1,2}^{ab}a_b^\mu)
\frac{\delta}{\delta \partial_\mu(\Omega_{1,2}^{ac}{\cal A}_c^\mu(u))}
\; ,
\end{equation}
with $\Omega_{1,2}$ defined as in \eq\nr{eq:wilson-def} with
$y\equiv\ln(P^\pm/\Lambda^\pm)$.

The next step in evaluating the NLO corrections is to integrate over
the momentum $\k$ in \eq(\ref{eq:O-NLO}). One integral appears
explicitly in the term that has two operators ${\mathbbm T}$, and
another momentum integral is hidden in the source term of the
fluctuation field $\beta^\mu$. Since in the CGC effective theory the
modes described as fields are bounded by $k^\pm<\Lambda^\pm$, these
longitudinal momentum integrals have an upper bound.  We shall compute
only the contribution of modes in the small slices
$\Lambda^{\prime+}<k^+<\Lambda^+$ and
$\Lambda^{\prime-}<k^-<\Lambda^-$.  By integrating out the field modes
in these slices, one is going from the original CGC EFT to a new
CGC${}^\prime$ EFT. The latter differs from the former because it has
an additional layer of (slower) sources while its field modes now
extend only up to smaller cutoffs $\Lambda^{\prime\pm}$.

Using the results of paper I, we have to leading logarithmic accuracy,
\begin{eqnarray}
&&
\int\limits_{\Sigma}\ud^3\vec\u\,
\big[\beta\cdot{\mathbbm T}_\u\big] 
+
\frac{1}{2}\sum_{\lambda,a}\int\frac{\ud^3\k}{(2\pi)^32E_\k}
\int\limits_{\Sigma} \ud^3\vec\u\, \ud^3\vec\v\,
\big[a_{-\k\lambda a}\cdot{\mathbbm T}_\u\big]
\big[a_{+\k\lambda a}\cdot{\mathbbm T}_\v\big]
\nonumber\\
&&\qquad\qquad
\empile{=}\over{\Lambda^{\prime\pm}<k^\pm< \Lambda^\pm}\quad
\ln\left(\frac{\Lambda^+}{\Lambda^{\prime+}}\right)\,{\cal H}_{\Lambda^+}
+
\ln\left(\frac{\Lambda^-}{\Lambda^{\prime-}}\right)\,{\cal H}_{\Lambda^-}\; .
\label{eq:JIMWLK}
\end{eqnarray}
In this equation, ${\cal H}_{\Lambda^\pm}$ are the JIMWLK Hamiltonians
of the right and left moving nuclei respectively, at the scales
$\Lambda^\pm$.  For the nucleus moving in the $+z$ direction, the
explicit form of the JIMWLK Hamiltonian is
\begin{equation}
{\cal H}_{\Lambda^+}
\equiv \frac{1}{2}
\!\!\!\int\limits_{\xt,\yt}\!\!\!
\frac{\delta}{\delta\widetilde{\mathcal{A}}_a^+(\epsilon^-,\yt)}
\eta^{ab}_1(\xt,\yt)
\frac{\delta}{\delta\widetilde{\mathcal{A}}_b^+(\epsilon^-,\xt)}, 
\end{equation}
with $\epsilon^-=1/\Lambda^+$ and where\footnote{Wilson lines without
  a rapidity argument are defined as in \eq\nr{eq:wilson-def} with
  $y\equiv\ln(P^\pm/\Lambda^\pm)$ -- they integrate over all the
  sources of the CGC EFT down to the cutoff $\Lambda^\pm$.
  Moreover, the derivatives in the JIMWLK Hamiltonian are with respect
  to the slowest sources of the EFT.}
\begin{eqnarray}
&&
\!\!\!\!\!
\eta^{ab}_1(\xt,\yt)
=
\frac{1}{\pi}
\int \frac{d^2\u_\perp}{(2\pi)^2}
\;
\frac{(\x_\perp^i-\u_\perp^i)(\y_\perp^i-\u_\perp^i)}{(\x_\perp-\u_\perp)^2(\y_\perp-\u_\perp)^2}
\nonumber\\
&&
\times
\Big[1+
\Omega_1(\xt)\Omega_1^\dagger(\yt)
\!-\!\Omega_1(\xt)\Omega_1^\dagger(\ut)
\!-\!\Omega_1(\ut)\Omega_1^\dagger(\yt)
\Big]_{ab}.
\label{eq:eta-f}
\end{eqnarray}
There is a similar definition for the second nucleus moving in the
$-z$ direction.

An important point to note here is that the relation in
eq.~(\ref{eq:JIMWLK}) is a property of the operator enclosed in the
curly brackets of eq.~(\ref{eq:O-NLO}), regardless of the details of
the observable ${\cal O}$ under consideration.  The second remarkable
property of this result is that the leading logarithms can be assigned
to one or the other of the two nuclei, without any mixing that would
violate factorization.

The expectation value of ${\cal O}$, at NLO, can be represented in the
CGC effective theory as
\begin{eqnarray}
& &\Big<{\cal O}_{_{\rm LO}} + {\cal O}_{_{\rm NLO}}\Big> = 
\int 
\big[D\Omega_1(y,\xt)D\Omega_2(y,\xt)\big]
 \nonumber \\
&&
\qquad\quad\times
W_{_{\Lambda^+}}\big[\Omega_1(y,\xt)\big]
W_{_{\Lambda^-}}\big[\Omega_2(y,\xt)\big]\,
\big[ 
{\cal O}_{_{\rm LO}} + {\cal O}_{_{\rm NLO}} 
\big] ,
\label{eq:exp-value}
\end{eqnarray}
where $W_{_{\Lambda^\pm}}[\Omega_{1,2}(y,\x_\perp)]$ are the functional 
probability distributions for the Wilson line configurations
$\Omega_{1,2}(y,\x_\perp)$.

Inserting \eq(\ref{eq:JIMWLK}) in \eq(\ref{eq:O-NLO}) and substituting
the resulting expression on the r.h.s. of \eq(\ref{eq:exp-value}), one
can perform an integration by parts~\footnote{The JIMWLK Hamiltonian
  is Hermitian.} such that ${\cal H}_{\Lambda^\pm}$ operates on the
distributions $W_{_{\Lambda^\pm}}$. Let us denote\footnote{An
  identical analysis also applies to the second nucleus.}
\begin{equation}
W_{_{\Lambda^{\prime+}}}[\Omega^\prime_1(y,\x_\perp)\big]
\equiv
\Big[1\!+\!
\underbrace{\ln\left(\frac{\Lambda^+}{\Lambda^{\prime+}}\right)}_{\ud y}
{\cal H}_{\Lambda^+}\Big]
W_{_{\Lambda^+}}\big[\Omega_1(y,\x_\perp)\big]
\, .
\label{eq:JIMWLK-evol}
\end{equation}
This equation is the infinitesimal form of the JIMWLK evolution
equation, where $\Omega^\prime_1$ is the Wilson line corresponding to
a source distribution $\wt{\rho}^\prime_{1} = \wt{\rho}_{1} +
\delta{\wt{\rho}_{1}}$ and $\delta{\wt{\rho}_{1}}$ has support only in
the interval $[y,y+dy]$.

The argument of $W_{_{\Lambda^{\prime+}}}$ in
\eq(\ref{eq:JIMWLK-evol}) extends one step further in rapidity than
the argument of $W_{_{\Lambda^+}}$ -- it is defined over the range
$0\le y\le \ln(P^+/\Lambda^{\prime+})$ and is hence a Wilson line in
the CGC${}^\prime$ EFT. Thus our derivation proves that
\begin{equation}
\Big<{\cal O}_{_{\rm LO}}
+
\underbrace{{\cal O}_{_{\rm NLO}}}_{\Lambda^{\prime\pm}<k^\pm<\Lambda^\pm}
\Big>
=
\Big<{\cal O}_{_{\rm LO}}\Big>^\prime\; .
\label{eq:RG}
\end{equation}
The prime on the r.h.s. indicates that the average is performed with a
probability distribution corresponding to the Wilson lines of the
CGC${}^\prime$ effective theory.  In other words, this identity states
that the classical expectation value of ${\cal O}$ in the original
EFT, corrected by quantum fluctuations in a small slice of field
modes, can be expressed as a purely classical expectation value in a
new EFT with a lower cutoff and with a distribution of Wilson lines
evolved according to \eq(\ref{eq:JIMWLK-evol}).

Equation~(\ref{eq:RG}) describes how to resum the leading logarithmic
quantum corrections in a small slice of longitudinal momentum.
Successive leading logarithmic contributions down to $k^\pm=0$ are
obtained by repeating this elementary step infinitely many times while
letting the thickness of the slices go to zero. One then obtains
\begin{eqnarray}
\Big<{\cal O}\Big>_{_{\rm LLog}}
=
\int 
\big[D\Omega_1(y,\xt)D\Omega_2(y,\xt)\big]\;
W\big[\Omega_1(y,\xt)\big]W\big[\Omega_2(y,\xt)\big]\;
{\cal O}_{_{\rm LO}}\; ,
\label{eq:fact-gen}
\end{eqnarray}
where $W \equiv\lim_{\Lambda^\pm \to 0} W_{_{\Lambda^\pm}}$.  This
expression is the central result of this paper.
Eq.~(\ref{eq:fact-gen}) shows that all the leading logarithms of
rapidity, whether they correspond to the rapidity intervals between
the nuclei and the tagged gluons or between the various tagged gluons,
can be absorbed into the probability distributions $W$ for the
trajectories of Wilson lines of the two projectiles. {\sl This formula
  applies to any inclusive observable for which \eq(\ref{eq:O-NLO}) is
  valid, regardless of whether the observable is local in rapidity or
  not.}  Because our result contains an average over $y$-dependent
``trajectories'' of Wilson lines, rather than an average over Wilson
lines at a given fixed rapidity, it contains a lot of information
about multigluon correlations at different rapidities.

\section{One- and two-gluon inclusive spectra}
\label{sec:1and2gluon-spectra}
\begin{figure}[htbp]
\begin{center}
\resizebox*{!}{7cm}{\includegraphics{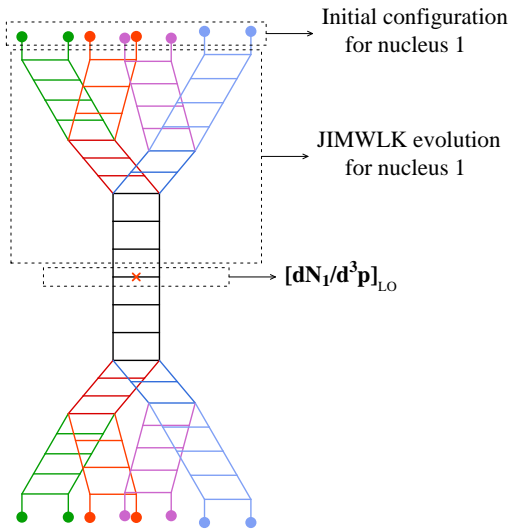}}
\end{center}
\caption{\label{fig:1-gluon} Diagrammatic representation of the
  various building blocks in the factorized formula for the inclusive
  single gluon spectrum. The lower part of the figure, representing nucleus 2, is made up of identical 
  building blocks.}
\end{figure}
We will now extract from our general result in \eq(\ref{eq:fact-gen}),
expressions for single and double inclusive gluon spectra.  The single
inclusive gluon spectrum $dN_1/\ud^3\p$ at LO depends only on the
Wilson lines $\Omega_{1,2}(y,\xt)$ at the rapidity $y=y_p$ of the
produced gluon and not on the whole rapidity range as in
\eq(\ref{eq:fact-gen}). Therefore, we can simplify
eq.~(\ref{eq:fact-gen}) by inserting the identity
 \begin{equation}
 1=\int \big[DU_{1,2}(\xt)\big]\;
 \delta\big[U_{1,2}(\xt)-\Omega_{1,2}(y_p,\xt)\big]
\label{eq:identity}
 \end{equation} 
 and by defining the corresponding probability distributions for
 configurations of Wilson lines at the rapidity $y_p$
\begin{equation}
Z_{y_p}[U_{1,2}(\xt)]
\equiv
\int\big[D\Omega_{1,2}(y,\xt)\big]\;
W\big[\Omega_{1,2}(y,\xt)\big]
\;
\delta\big[U_{1,2}(\xt)-\Omega_{1,2}(y_p,\xt)\big]\; .
\label{eq:Z}
\end{equation}
One then obtains the all order leading log result for the single
inclusive gluon spectrum in the following form
\begin{equation}
\left.
\Big<\frac{\ud N_1}{\ud^2\p_\perp dy_p}\Big>
\right|_{_{\rm LLog}}
=
\int
\big[D U_{1}(\xt)\,DU_{2}(\xt)\big] \;
Z_{y_p}\left[U_{1}\right]\,
Z_{y_p}\left[U_{2}\right]\;
\left.
\frac{\ud N_1\big[U_{1},U_{2}\big]}{\ud^2\p_\perp \ud y_p}
\right|_{_{\rm LO}}\!.
\label{eq:N1-resummed}
\end{equation}
Note that the distribution $Z_{y_p}[U]$ obeys the JIMWLK equation,
\begin{equation}
\partial_{y_p} Z_{y_p}[U] = {\cal H}_{y_p}\,Z_{y_p}[U]\; ,
\end{equation}
which must be supplemented by an initial condition at a rapidity
close to the fragmentation region of the projectiles.
Eq.~(\ref{eq:N1-resummed}) is illustrated in the figure
\ref{fig:1-gluon}.

At Leading Order, the inclusive two-gluon spectrum is simply the
disconnected product of two single gluon
spectra~\cite{Gelis:2008ad,Kharzeev:2004bw} each of which depends on
Wilson lines at the rapidity of the corresponding gluon. Using
eq.~(\ref{eq:identity}) in \eq(\ref{eq:fact-gen}) (now we need to
insert four such delta functions), one obtains for the resummed
two-gluon spectrum the expression
\begin{eqnarray}
&&
\left.
\frac{\ud N_2}{\ud^2 \pt \ud y_p  \ud^2 \qt \ud y_q}
\right|_{_{\rm LLog}}
=
\int
\big[D U^p_{1}(\xt)D U^p_{2}(\xt)D U^q_{1}(\xt)D U^q_{2}(\xt)\big] 
\nonumber\\
&&
\;\times
Z_{y_p,y_q}\big[U^p_{1},U^q_{1}\big]\,
Z_{y_p,y_q}\big[U^p_{2},U^q_{2}\big]\;
\left.
\frac{\ud N_1\big[U_{1}^p,U_{2}^p\big]}{\ud^2\p_\perp \ud y_p}
\right|_{_{\rm LO}}
\left.
\frac{\ud N_1\big[U^q_{1},U^q_{2}\big]}{\ud^2\q_\perp \ud y_q}
\right|_{_{\rm LO}}
,
\label{eq:N2-resummed-largey-1}
\end{eqnarray}
where we have introduced the double probability distribution of
Wilson lines as
\begin{eqnarray}
&&
Z_{y_p,y_q}\big[U^p_{1,2},U^q_{1,2}\big]
\equiv
\!\int\!\big[D\Omega_{1,2}(y,\xt)\big]\;
W\big[\Omega_{1,2}(y,\xt)\big]
\nonumber\\
&&
\qquad
\times
\delta\big[U^p_{1,2}(\xt)\!\!-\!\Omega_{1,2}(y_p,\xt)\big]
\delta\big[U^q_{1,2}(\xt)\!\!-\!\Omega_{1,2}(y_q,\xt)\big]
\, .
\nonumber\\
&&
\end{eqnarray}
This double distribution obeys the JIMWLK equation with respect to the
largest of the two rapidities,
\begin{equation}
\mbox{if\ }y_q>y_p,\quad
\partial_{y_q}
Z_{y_p,y_q}\big[U^p,U^q\big]
=
{\cal H}_{y_q}\,Z_{y_p,y_q}\big[U^p,U^q\big]\; ,
\end{equation}
with the boundary condition
\begin{equation}
\lim_{y_q\to y_p} Z_{y_p,y_q}\big[U^p,U^q\big]
=
Z_{y_p}\big[U^p\big]\,\delta\big[U^p-U^q\big]\; .
\end{equation}
Alternately, this double distribution can be expressed as 
\begin{equation} \label{eq:markov}
Z_{y_p,y_q}\big[U^p,U^q\big] 
=
G_{y_q,y_p}\big[U^q,U^p\big] 
Z_{y_p}\big[U^p\big],
\end{equation}
where the Green's function $G_{y_q,y_p}\big[U^q,U^p\big] $ satisfies
the JIMWLK equation\footnote{We use here, and previously, the fact
  that the JIMWLK Hamiltonian acts only on objects at equal or higher
  rapidities.}
\begin{equation}
\partial_{y_q} G_{y_q,y_p}\big[U^q,U^p\big] 
= \mathcal{H}_{y_q}\,G_{y_q,y_p}\big[U^q,U^p\big]\; ,
\label{eq:green-JIMWLK}
\end{equation}
with the initial condition
\begin{equation} \label{eq:greenic}
\lim_{y_q \to y_p} G_{y_q,y_p}\big[U^q,U^p\big] 
=
\delta\big[U^q-U^q\big]\; .
\end{equation}
This Green's function describes multigluon evolution, between two
tagged gluons, in the presence of strong color sources from the
projectiles.

\begin{figure}[htbp]
\begin{center}
\resizebox*{!}{7cm}{\includegraphics{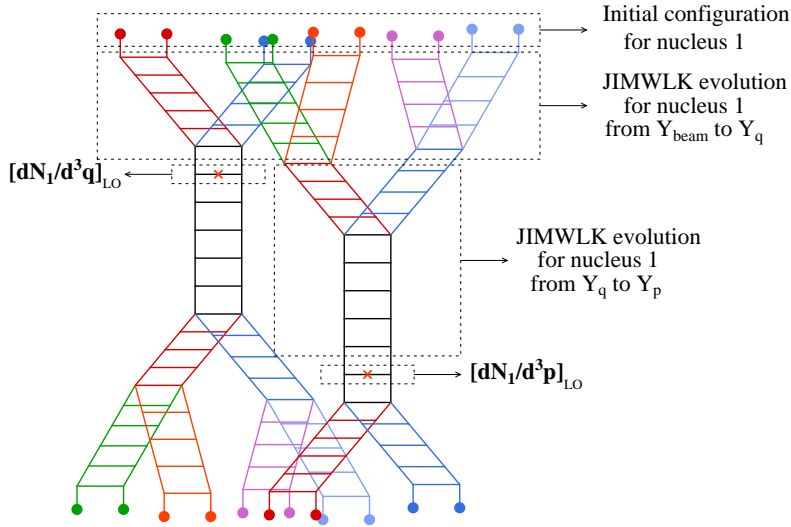}}
\end{center}
\caption{\label{fig:2-gluon} Diagrammatic representation of the
  various building blocks in the factorized formula for the inclusive
  2-gluon spectrum. As in the previous figure, the corresponding evolution from nucleus 2 at the bottom of the 
  figure is not shown explicitly.}
\end{figure}

Our result for the double inclusive gluon spectrum, to leading
logarithmic accuracy, can thus be expressed as follows
\begin{eqnarray}
&&
\left.
\Big<\frac{\ud N_2}{\ud^2 \pt \ud y_p  \ud^2 \qt \ud y_q}\Big>
\right|_{_{\rm LLog}}
=
\int
\big[D U^p_{1}(\xt)D U^p_{2}(\xt)D U^q_{1}(\xt)D U^q_{2}(\xt)\big] 
\nonumber\\
&&
\qquad\qquad\qquad
\times
Z_{y_p}\big[U^p_{1}\big]\,G_{y_p,y_q}\big[U^p_{1},U^q_{1}\big]
Z_{y_q}\big[U^q_{2}\big]\,G_{y_q,y_p}\big[U^q_{2},U^p_{2}\big]
\nonumber\\
&&
\qquad\qquad\qquad
\times
\left.
\frac{\ud N_1\big[U_{1}^p,U_{2}^p\big]}{\ud^2\p_\perp \ud y_p}
\right|_{_{\rm LO}}
\left.
\frac{\ud N_1\big[U^q_{1},U^q_{2}\big]}{\ud^2\q_\perp \ud y_q}
\right|_{_{\rm LO}}
\; .
\label{eq:N2-resummed-largey}
\end{eqnarray}
Eq.~(\ref{eq:N2-resummed-largey}) generalizes the result in paper II
-- that result, as implied by \eq(\ref{eq:greenic}), is recovered when
the rapidities of the two gluons are close to each other. Our formula
for the two-gluon spectrum in eq.~(\ref{eq:N2-resummed-largey}) is
illustrated in fig.~(\ref{fig:2-gluon}).  By using
\eq(\ref{eq:fact-gen}), it is straightforward to write down similar
formulae for higher gluon correlations.

Factorization is obviously manifest in eq.~(\ref{eq:fact-gen}).  It is
slightly less apparent in eq.~(\ref{eq:N2-resummed-largey}) because
there are more factors in the integrand. However, both the
$Z$-functionals and the $G$ Green's functionals are universal objects
that describe the partonic content of a nucleus at high energy. The
part of the integrand which is specific to the observable under
consideration is relegated to the factors $dN_1/d^3\p$ and $dN_1/d^3\q$.

Note also that these distributions are far more general than the
$k_\perp$--de\-pen\-dent correlators of Wilson lines that are often
discussed in the literature of high energy QCD. The latter appear in a
form of factorization which goes under the rubric of
$k_\perp$-factorization. This type of factorization is formulated in
terms of single gluon distributions, but is known to be
violated~\cite{JalilianMarian:2004da} for 2-gluon correlations in
collisions involving at least one saturated projectile. It is in this
context that one should interpret the results of~\cite{oldstuff} which
concludes that factorization is violated. In stark contrast, our
universal density functionals and Green's functions contain all the
relevant information on rapidity dependent $n$-gluon correlations. In
fact, our factorization result is a general consequence of causality
and for this reason should even be valid beyond leading log accuracy.

In practice, solving the JIMWLK equation to compute the r.h.s of
\eq(\ref{eq:N2-resummed-largey}) is more conveniently achieved by
writing this equation as a Langevin equation for Wilson lines living
on the SU(3) group manifold~\cite{Blaizot:2002xy}.  This stochastic
formulation was implemented in the only extant numerical study of the
JIMWLK equation~\cite{Rummukainen:2003ns}.  Because solving the JIMWLK
equation can be numerically challenging, a simpler formulation of high
energy evolution is provided by the Balitsky--Kovchegov (BK)
equation~\cite{BK}, which is a closed form mean field simplification
of the JIMWLK expression for 2-point Wilson line correlators in the
CGC~\footnote{The BK equation is valid in the large $\nc$ limit for
  large nuclei.}. The BK equation corresponds to a nonlocal Gaussian
form of the $Z_{y_p},Z_{y_q}$ functionals~\cite{Fujii:2006ab} (when
expressed in terms of the color source distributions) with a variance
$\mu_A^2(y_{p,q},x_\perp)$.  Because the weight functionals at both
$y_p$ and $y_q$ have this Gaussian form, the Green's function
$G_{y_p,y_q}$ connecting the two must also be a Gaussian, whose
variance can be determined from numerical solutions of the BK
equation. Therefore, quantitative results for
\eq(\ref{eq:N2-resummed-largey}) can be obtained within this BK
framework. These will be discussed in future work.

\section{Dilute-dense limit}
\label{sec:dilute}
The results obtained here are valid for the collision of two dense
projectiles whose color charge densities of both are given by
$\wt\rho_{1,2}\sim g^{-1}$. This is the case in large nuclei or in
nucleons at very high energies. An interesting question is whether
decreasing the magnitude of $\wt\rho_1$ and/or $\wt\rho_2$ in the
formulas we have obtained so far gives the correct answer for
dilute-dense or dilute-dilute collisions. Before going further, let us
note that the permitted range for dilute to dense color charge
densities is between $g$ and $g^{-1}$. The upper value, assumed in our
study, corresponds to a fully saturated projectile. The lower value
corresponds to a very dilute projectile whose parton density is of
order unity ensuring that its density of color charge is proportional
to $g$.

The answer to the question posed is affirmative for the single
inclusive gluon spectrum (eq.~(\ref{eq:N1-resummed})). By taking the
limit $\wt\rho_2\sim g$ in this formula, one recovers immediately the
well known result for the single inclusive spectrum in pA collisions
and likewise for $pp$ collisions when we let both $\wt\rho_{1,2}$
become of order $g$.

\begin{figure}[htbp]
\begin{center}
\resizebox*{!}{7cm}{\includegraphics{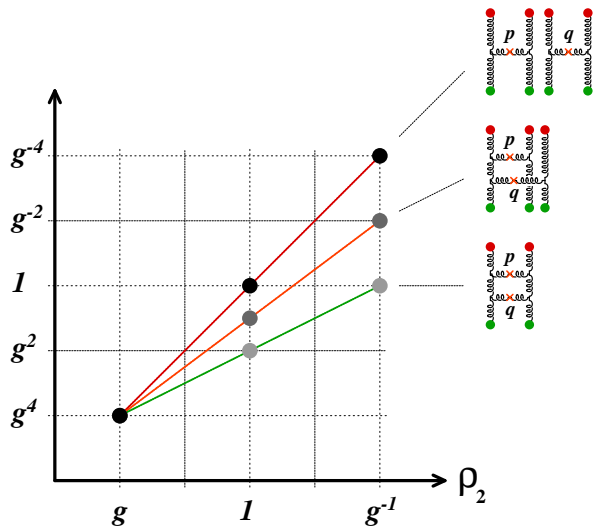}}
\end{center}
\caption{\label{fig:dilute}Order of magnitude of various contributions
  to the 2-gluon spectrum as a function of the color charge density
  $\rho_2$ in the small projectile (the color charge density in the
  large projectile is held fixed $\rho_1={\cal O}(g^{-1})$).  Large
  logarithms of the energy, which become relevant in the leading
  logarithmic resummation, are not considered here.}
\end{figure}

However, taking this limit in the 2-gluon spectrum
(eq.~(\ref{eq:N2-resummed-largey})) does not lead to the correct
results for the inclusive 2-gluon spectrum in pA or pp
collisions~\cite{JalilianMarian:2004da}. The reason of this
discrepancy is that the corresponding power counting for the two gluon
spectrum is very different for ``dense'' color sources $\sim g^{-1}$
relative to the case of ``dilute'' sources $\sim g$. In the power
counting for dense sources $\sim g^{-1}$, certain graphs are
suppressed that would also be leading graphs for dilute sources $\sim
g$. This is illustrated in figure \ref{fig:dilute}, where we have
displayed the order of magnitude of three different contributions to
the 2-gluon spectrum, as a function of $\wt\rho_2$. (For pA
collisions, $\wt\rho_1$ is fixed to be of order $g^{-1}$.) One sees
that in the dense case, only one of these graphs is important, while
they are all important in the dilute limit. Since
eq.~(\ref{eq:N2-resummed-largey}) is obtained by assuming dense
projectiles, it contains only the disconnected graph and misses the
other two -- this implies that eq.~(\ref{eq:N2-resummed-largey}) is
not the complete answer in the dilute limit.  As noted, this subtlety
affects correlations between two or more gluons but not the single
gluon spectrum.

The previous discussion only deals with contributions at Leading
Order. However, a similar discrepancy between the power counting in
the dense-dense and dilute-dense limits occurs in the resummation of
the leading logarithmic contributions. There, one sees that the
operators included in the JIMWLK Hamiltonian are not the only ones
that contribute at leading log accuracy in the dilute regime. For
instance, when $\wt\rho\sim g$, an operator of the form $g^4 \wt\rho^2
(\delta/\delta\wt\rho)^4$ has the same order of magnitude as the
operators in the JIMWLK Hamiltonian in the dilute limit (but is
totally suppressed in the dense regime). Such an operator, with a
prefactor of order $\wt\rho^2$ and four derivatives with respect to
the color source, corresponds to ``pomeron splittings'' in the energy
evolution~\cite{JalilianMarian:2004da,Mueller:2005ut,Iancu:2005nj,Kovner:2005nq,Kovner:2005uw}
-- while the JIMWLK evolution only has ``pomeron mergings'' (because
the number of $\wt\rho$'s in the prefactor is always equal to or
greater than the number of derivatives). In principle, one would like
a formalism where both limits contain the right physics. Unlike
previous works which address the full S-matrix for high energy
scattering, our focus will be on the more limited goal of computing
inclusive gluon spectra in dense-dilute collisions. We believe that
substantial progress in this direction is feasible and will further
address this topic in a future publication.

\section{Conclusions}

In this paper, we obtained in eq.~(\ref{eq:fact-gen}) a general result
for inclusive n-gluon production at arbitrary rapidities in the
collision of two dense projectiles (such as heavy nuclei) with charge
densities given by $\wt\rho_{1,2}\sim g^{-1}$. The result is expressed
in terms of universal $W$-density matrix functionals which contain
information on n-gluon correlations in the wavefunctions of the dense
projectiles.  Our formalism is strictly valid in the leading
logarithmic approximation in $x$. We anticipate, on the basis of
simple causality arguments, that the structure of our result will hold
beyond leading logarithmic accuracy.

We explicitly wrote down the corresponding expressions for single and
double inclusive gluon production with arbitrarily large rapidity
separations between tagged gluons. Until this point, there was no
microscopic QCD based formalism that allows the computation of the
near side ridge correlations in nucleus-nucleus collisions {\it when
  the rapidity separation between the measured particles is of the
  order of $1/\alpha_s$ or more.} Our formalism fills this gap and
allows for future quantitative comparisons and predictions for the
rapidity dependence of the the ridge like structures observed in
central nucleus--nucleus collisions at RHIC and in future at the LHC.
At the LHC, one may have the possibility of studying such structures
that may span $6$-$10$ units in rapidity. Such long range correlations
therefore open a new window on the study of multiparton correlations
in QCD as well as the provide a ``chronometer'' of the strong field
initial ``Glasma'' stage of heavy ion collisions.

When both projectiles are dense, we argued that only ``Pomeron
mergings'' that are fully included in the JIMWLK Hamiltonian are
relevant for inclusive gluon production. When one or both of the
projectiles becomes dilute, our power counting suggests that ``Pomeron
splitting'' contributions, become equally important as the ``merging''
contributions for correlations involving two or more gluons. Because
these are not included in the JIMWLK formalism, they cannot be
obtained by taking a naive low density limit of the dense-dense
formalism discussed in this paper. A smooth interpolation from the
dilute-dense to the dense-dense limits for multigluon inclusive
distributions requires that we first compute corrections to the JIMWLK
Hamiltonian in the dilute-dense limit. While there have been several
such studies in the context of the S-matrix for high energy
scattering, they are in their infancy for inclusive multigluon
production~\cite{Kovner:2006wr}. These studies will be important for
extending our studies for nucleus--nucleus collisions to asymmetric
systems such as high energy proton--nucleus collisions. This work is
in progress and will be addressed in a future publication.

\section*{Acknowledgements}
We thank the Center for Theoretical Sciences of the Tata Institute for
Fundamental Research for their support during the program ``Initial
Conditions in Heavy Ion Collisions''. R.V.'s research is supported by
the US Department of Energy under DOE Contract No.  DE-AC02-98CH10886.
F.G.'s work is supported in part by Agence Nationale de la Recherche
via the programme ANR-06-BLAN-0285-01.

\end{document}